# Supervised Machine Learning Techniques: An Overview with Applications to Banking

Linwei Hu, Jie Chen, Joel Vaughan, Hanyu Yang, Kelly Wang, Agus Sudjianto, and Vijayan N. Nair[1]

Corporate Model Risk, Wells Fargo

July 26, 2020

## Abstract

This article provides an overview of Supervised Machine Learning (SML) with a focus on applications to banking. The SML techniques covered include Bagging (Random Forest or RF), Boosting (Gradient Boosting Machine or GBM) and Neural Networks (NNs). We begin with an introduction to ML tasks and techniques. This is followed by a description of: i) tree-based ensemble algorithms including Bagging with RF and Boosting with GBMs, ii) Feedforward NNs, iii) a discussion of hyper-parameter optimization techniques, and iv) machine learning interpretability. The paper concludes with a comparison of the features of different ML algorithms. Examples taken from credit risk modeling in banking are used throughout the paper to illustrate the techniques and interpret the results of the algorithms.

## 1 Machine Learning

### 1.1 Introduction

Although the history of Machine Learning (ML) dates back to at least the 1950s, the techniques have seen wide usage only in the last two to three decades. The main reasons are recent advances in computing power, increasing availability of open-source software, and developments in data capture and database technologies. All of these have led to the ability to collect, manipulate and analyze massive datasets.

The term machine learning was coined by Arthur Samuel in 1959 at IBM who called it:

i. *"A field of study that gives computers the ability to learn without being explicitly programmed."*

A more engineering-oriented definition was given by Tom Mitchell in 1997:

ii. *A computer program is said to learn from experience E with respect to some task T and some performance measure P, if its performance on T, as measured by P, improves with experience E.*

Consider, for example, a spam filtering algorithm used in email boxes as an ML algorithm. The task T is to tag if an email is a spam or not, the experience E is the training data, and the performance measure is the accuracy. As we get more data, we can train the algorithm to learn more spam patterns and get better in tagging spam vs non-spam emails.

There is some confusion in the usage of the terms *Artificial Intelligence* and *Machine Learning*, with some people viewing them to be interchangeable. AI is a much broader field, and ML is a subset and one of the pathways to accomplishing the goals of an AI project. The term AI was first coined by John McCarthy in 1959 as:

iii. *The study of "intelligent agents" – devices that perceive the environment and take actions that maximize its chance of success at some goal.*

However, AI has a very long history dating back to formal reasoning in logic, philosophy and other fields. It has had mixed success in the past and has gone through periods of "AI winters". But it has had a massive resurgence in recent years, again due to the amounts of data available and exponential leaps in computing power. In particular, the development of deep learning neural networks with their excellent predictive performance, especially in the area of pattern recognition, have led to much excitement.

---

[1] Corresponding author (e-mail: Vijayan.Nair@wellsfargo.com)

Coming back to ML, the algorithms are typically classified into three primary groups:

1) <u>Supervised Learning</u>: Most common examples are classification and regression problems where there is a "label"– a dependent variable $Y$, that is a continuous response or class membership. This allows the algorithm to learn from the training data and develop predictions or classifications for new observations. $Y$ can be binary or categorical responses in classification problems, numerical responses in general regression problems, or a mixture of the two. Regression algorithms for binary data can be used for classification by thresholding the numerical predicted scores at an appropriate value. There are variations of supervised learning such as semi-supervised where the training set consists of a mixture of labeled and unlabeled data.

2) <u>Unsupervised Learning</u>: There are no dependent variables in this case, and the goal is to discover relevant patterns or the underlying structure in a given set of (often high-dimensional) independent variables. Common examples include dimension reduction, clustering, and anomaly detection. Data mining techniques such as pairwise association algorithms also belong to this class.

3) <u>Reinforcement Learning (RL):</u> This is quite different from the first two classes of problems. RL involves "goal-oriented learning" based on a decision space, an action space, and a reward structure. It uses various techniques to explore the space, move towards optimal solutions, and make decisions. RL offers a lot of promise in pushing the boundaries of AI, and it has already been used in applications such as personalized medicine, Google's alpha-go, and others.

There are also newer categories of learning problems that are being introduced over time such as Representation Learning, Transfer Learning, etc. In this article, we shall focus on Supervised Machine Learning (SML).

## 1.2  SML: Additional Background

There are multiple reasons for the increased interest in ML algorithms in general and SML in particular. Here are some selected ones:

a) When there is a huge amount of data, there is no need to restrict attention to parametric modeling techniques that are limited in their ability to capture complex input-output relationships. This has been recognized within the traditional statistical community, where there has been a huge impetus to develop more flexible techniques. There has been increased use of nonparametric and semiparametric methods, decision trees (including classification and regression trees), projection pursuit or additive-index models, sliced inverse regression, flexible principal components analysis, and so on.

b) While the SML techniques discussed here can be viewed as part of flexible nonparametric regression and classification techniques, one big difference is the emphasis in automating many of the steps in model building. Traditional statistical techniques require a considerable amount of manual effort for variable selection and feature engineering, tasks which are mostly automated in SML.

c) A third reason has been captured well by the paper "Statistical Modeling: The Two Cultures" by Breiman [1]. It contrasts the dual goals of prediction vs model fitting – the former based on so-called algorithmic modeling (using the observed data and relying on test/validation data to calibrate the results) while model-fitting emphasizes understanding the data generative process and interpretation of input-output functional relationships, estimated regression coefficients, and traditional inference. Breiman argues in [1] that "*The statistical community has been committed to the almost exclusive use* *of data models [and model interpretation]. This commitment has led to irrelevant theory, questionable conclusions, and has kept statisticians from working on a large range of interesting current problems. Algorithmic modeling, both in theory and practice, has developed rapidly in fields outside statistics. It can be used both on large complex data sets and as a more accurate and informative alternative to data modeling on smaller data sets. If our goal as a field is to use data to solve problems, then we need to move away from exclusive dependence on data models and adopt a more diverse set of tools."* These comments and the main thesis of the paper created significant controversy in the statistical community (see discussions following the paper by several distinguished researchers).

If one were to focus solely on predictive performance, then the complex ML algorithms based on neural networks or ensemble methods (such as boosting or bagging) typically perform as well or better than even advanced statistical models. However, the fitted models are opaque, and it is not easy to unravel the input-output relationships.

In banking applications at least, we are interested in both goals: i) we want models that have good predictive performance subject to appropriate regularization so as not to over-fit; and ii) we also want to be able to understand the model, the input-output relationships, conformity to subject matter knowledge, possible problems caused by collinearity and so on. It is not easy to achieve both; in particular, it is challenging to understand the input-output relationships and interpret the results of ML models. We will discuss some of the approaches currently available in the literature and emerging research directions and guidelines.

The rest of the document is organized as follows. It describes tree-based ensemble algorithms including Bagging with Random Forest and Boosting with Gradient Boosting Machines. This is followed by a description of Feedforward Neural Networks. The next section deals with techniques for tuning (optimizing) hyper-parameters in these algorithms. Diagnostics for interpreting the ML algorithms are then discussed. The article ends with a comparison of SML algorithms and discussion of related topics.

## 2  Tree-Based SML Algorithms

### 2.1  Classification and Regression Tree (CART)

We begin with an overview of decision trees since they are the building blocks of the SML algorithms discussed in this section. There are many tree-based algorithms for classification and regression: Chi-squared Automatic Interaction Detector (CHAID) developed by Kass [2], the Ross Quinlan trilogy (ID3, C4.5 and C5.0) [3], CART (Classification and Regression Trees) developed by Breiman et al. [4], and others like conditional-inference trees [5]. CART is the most popular technique, is available in open-source software[2], and hence is our focus here.

CART recursively partitions the data into increasingly "homogeneous" groups based on some specified criterion until a stopping rule is reached. The algorithm works as follows:
1. Start from the root note with all the data.
2. Split each node into two child nodes to minimize some impurity measure (defined later). The best split is found by searching through all possible combinations of variables and their split points.
3. The tree is grown until a stopping rule is reached. Tree size is controlled by several hyper-parameters which are selected by hyper-parameter tuning.
4. Finally, data within each terminal node (or leaf) is used to prediction: node sample mean for continuous responses and majority vote or class proportions for binary/categorical responses.

---

[2] For R, rpart implements CART; for python, scikit learn's sklearn.tree module implements CART.

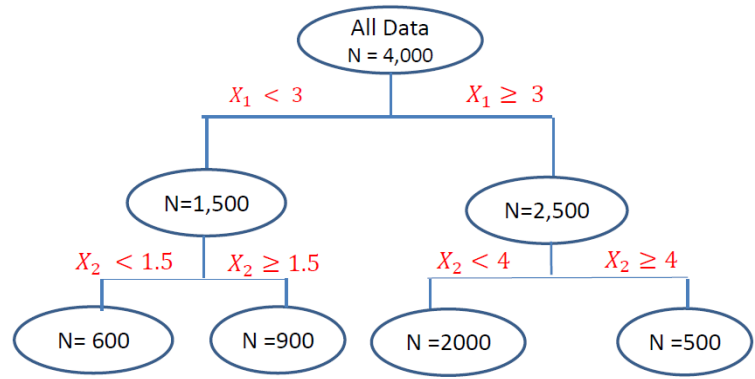

Figure 1. Illustrative regression tree with depth 2

Figure **1** shows a hypothetical regression tree with the root node comprising the entire dataset with 4,000 observations. In the first split, the observations are partitioned into two nodes based on the values of the predictor: $X_1 < 3$ or $\geq 3$. This results in 1,500 observations in the left node and 2,500 observations in the right one. In the next step, the left node is further split into two nodes based on $X_2 < 1.5$ or $\geq 1.5$. On the other branch, the right node with 2,500 observations is again split based on $X_2$: $X_2 < 4$ or $\geq 4$. The stopping rule of maximum depth = 2 is reached, so there is no further splitting. In summary, the original dataset has been partitioned into four datasets corresponding to: i) $\{X_1 < 3, X_2 < 1.5\}$ with N = 600 observations; ii) $\{X_1 < 3, X_2 \geq 1.5\}$ with N = 900 observations; iii) $\{X_1 \geq 3, X_2 < 4\}$ with N = 2,000 observations; and iv) $\{X_1 \geq 3, X_2 \geq 4\}$ with N = 500 observations. The results of the tree can be viewed as a piecewise constant regression model:

$$\hat{y} = c_1\ I\{X_1 < 3, X_2 < 1.5\} + c_2\ I\{X_1 < 3, X_2 \geq 1.5\} + c_3\ I\{X_1 \geq 3, X_2 < 4\} + c_4\ I\{X_1 \geq 3, X_2 \geq 4\},$$

for some estimated constants $c_j$, $j = 1, \ldots, 4$. While most software packages use binary splits, there are algorithms where $k-$nary splits for $k > 2$ are used (see for example, CHAID).

For continuous responses, the impurity of a node $m$ is typically measured by mean squared error. For categorical responses, common impurity measures are Gini index, $\sum_{k=1}^{K} p_{mk}(1 - p_{mk})$, or entropy measure, $-\sum_{k=1}^{K} p_{mk} \log(p_{mk})$, where $p_{mk}$ is the proportion of class $k$ observations in node $m$. The two latter metrics typically produce similar results.

A large tree can over-fit and a small tree will under-fit, so controlling the size of the tree is important. This is done by tuning the parameters associated with the tree structure (called *hyper-parameters*). They include maximum depth, maximum number of terminal nodes, minimum node size (the number of observations or total sample weight in the node), improvement on impurity, etc. Some of these are closely related, and there is no need to tune all of them. One approach recommended in the literature is to grow the tree to be relatively large and then prune back using some threshold on cost complexity (see, for example, Hastie et. al. [6]).

Tree algorithms have several advantages. They are fast to build, intuitive to interpret, are able to handle both numeric and categorical data, are robust to outliers in predictors, and model nonlinearity and interactions automatically. Some implementations can even handle missing values without imputation. On the other hand, trees can be unstable. A small change in the data can result in a vastly different tree structure, especially if the tree is deep or the predictors have high correlation. In addition, their predictive performance is not as good as some other models. Finally, predictions are piecewise constant and hence not smooth. This will degrade performance in regression problems where the underlying function is smooth.

## 2.2 Ensemble Methods

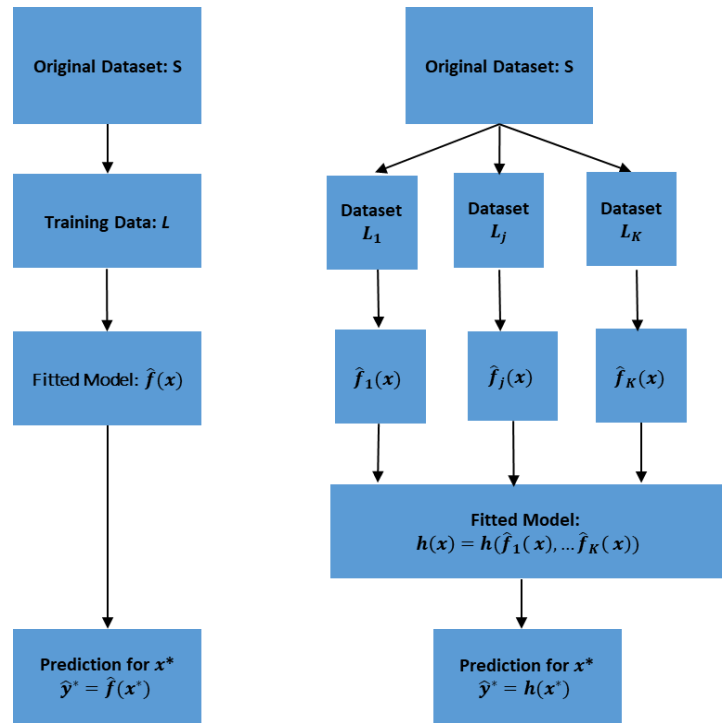

Figure 2: A visual comparison of traditional and ensemble techniques for supervised learning: left panel shows traditional approach while right panel shows an ensemble approach.

Ensemble learning is based on the idea of combining a set of *weak base learners*, like trees, to produce a more powerful algorithm. Figure 2 provides a visual display of ensemble methods for supervised learning and comparison with traditional regression. $S$ refers to the original dataset, $L_k$'s refers to the training datasets, and $\hat{f}_k$'s refer to the fitted models on the training data sets. In the traditional framework, we fit a single regression model and the results are easy to interpret. In ensemble models, the models are fitted across multiple training datasets and aggregated to get an overall model. While this leads to a flexible model that has very good predictive performance, the results are complex and difficult to understand/interpret.

### 2.2.1 Bagging and Random Forest

Bagging (bootstrap aggregating) was developed by Breiman and his collaborators [7] and is one example of an ensemble method. It works as follows:
1. Take many bootstrap samples;
2. Fit the base learner (such as a deep tree) to each bootstrap sample to get a base model; and
3. Combine all the base-model predictions by averaging (continuous response) or majority voting (categorical response).

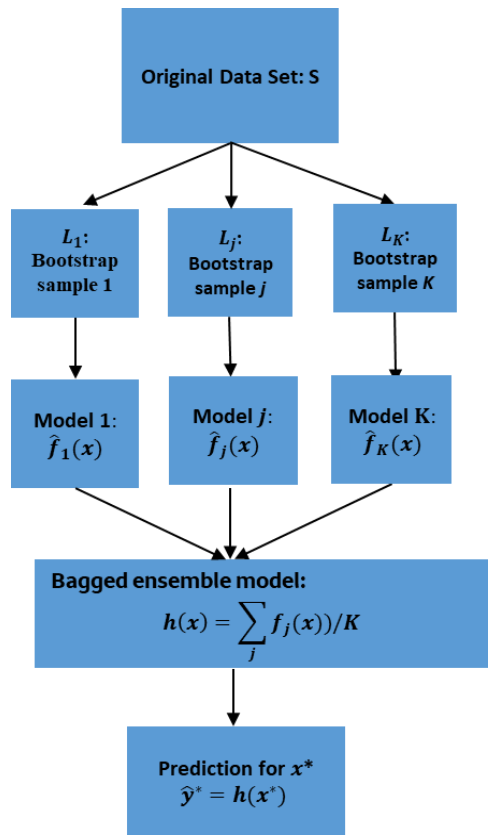

Figure 3. An illustration of Bagging for continuous response

Breiman [8] showed that bagging leads to "improvements for unstable procedures", such as deep tree algorithms because averaging reduces variance. The bootstrap samples have overlapping data in bagging methods, so the base model predictions are not independent, and consequently the variance of the average prediction will not converge to zero. The amount of variance reduction depends on the strength of correlation in the base predictions.

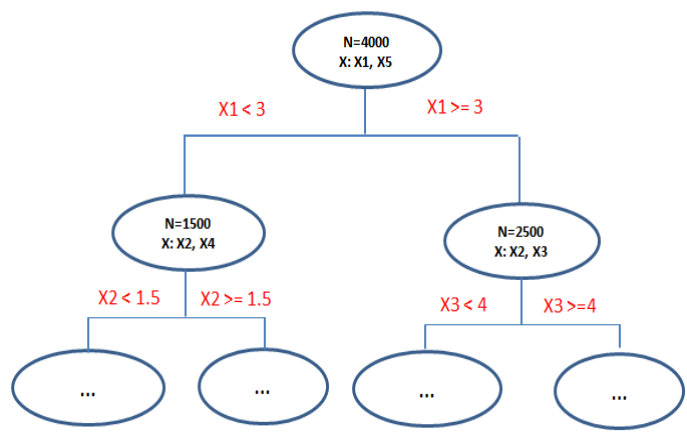

Figure 4. Random feature subset selection: randomly select 2 features for each split

Random forest (RF), the most well-known example of bagging algorithm, was popularized by Breiman [9]. It grows multiple trees by bootstrapping the observations (rows). In addition, to de-correlate the base model predictions to achieve a larger amount of variance reduction, each tree split is based on a random subset of the features (columns). An illustration is given in Figure 4 where two features are selected as split variables for each split, and the best split is only searched within these two variables.

To summarize, RF works as follows:
1. Take a bootstrap sample at each iteration;
2. Fit a deep tree algorithm to each bootstrap sample; at each tree split, restrict the set of features to a random subset; and
3. Combine all tree predictions by averaging (continuous response) or voting (categorical response).

There are several hyper-parameters associated with RF, such as number of trees and number of features to sample in each split. Deep trees are used in RF, so there is no need for pruning. In terms of the number of trees, in principle it is better to have as many as possible, but the predictions will stabilize after several hundred trees, after which additional trees do not substantially improve performance despite taking more time to run. Finally, for the number of features to sample in each split, it has been suggested in [6] that a value on the order of $\sqrt{p}$ works well for categorical responses and order of $p/3$ is good for continuous responses. Thus, there are not many hyper-parameters to tune for RF.

### 2.2.2 An example

We use an application on real estate mortgages to illustrate the results. The dataset had 15 million accounts, and the response was binary: whether the loan was "in trouble" or not. The term in-trouble is used to capture multiple categories: bankruptcy, short sale, 180+ days of delinquency in payments, etc. There were 54 predictors: macroeconomic variables (unemployment rate, GDP, and so on), static loan origination variables (FICO at origination time, loan interest rate, loan-to-value ratio, etc.), and time-varying loan characteristic variables (FICO over time, months in delinquency over time, etc.). We used a subset of the data for "prime real estate" segment for the discussion below.

We compare the results from a time-varying logistic regression model with those from RF first. For building the logistic regression model, the original team did variable selection and chose six variables: i) forecasted loan-to-value (LTV) ratio; ii) FICO over time; iii) indicator for before or after financial crisis; iv) unemployment rate; v) total personal income year over year ratio; and vi) delinquency status. They further grouped LTV and FICO into bins and fitted piecewise constant terms within bins to account for nonlinearity (this is a standard process in certain areas of modeling in banks). The RF algorithms used all 54 variables without manual transformations. The dataset was split into training, validation and test sets; the algorithms were developed on the training set; the tuning parameters optimized on validation set using grid search; and the final results were compared on the test set.

Figure 5 shows the ROC curves comparing logistic regression (red) with RF (blue) on hold-out test data. The green curve corresponds to GBM (Gradient Boosting) algorithm discussed in the next section. We see that RF is consistently better than logistic regression, and sometimes considerably so. The area under the curve (AUC), is 0.799 for logistic regression compared to 0.852 for RF – about a 9% increase (or `lift') in predictive performance, which is substantial in this type of applications. One could argue that this is an apples-to-oranges comparison since the logistic regression model is based on only six predictors while RF used all 54. But this is one of the major advantages of ML algorithms: one does not have to go through manual and time-consuming feature engineering (variable selection, identifying suitable transformations, and interactions). In practice, there may be several hundred variables and millions of observations, so manual feature engineering can be a formidable task.

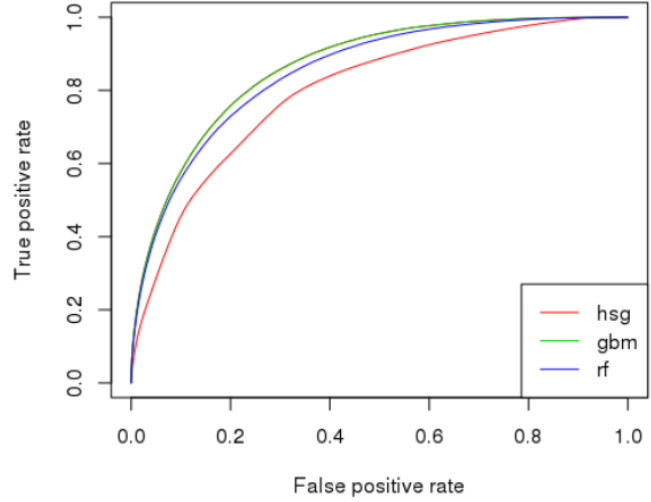

Figure 5: Comparison of ROC curves for three algorithms: logistic regression (hsg); RF; and GBM

### 2.2.3 Boosting

Boosting is another ensemble algorithm, one that is designed to repeatedly reduce errors in weak learners. Kearns [10] and Kearns and Valiant [11] posed the following question: "Can a weak learning algorithm be boosted into a stronger learning algorithm"? Schapire [12] answered this question in the affirmative and came up with the first boosting algorithm. But it had practical drawbacks (in terms of implementation) which were addressed by the AdaBoost algorithm introduced by Freund and Schapire [13].

Breiman [15] made the path-breaking observation that AdaBoost can be viewed as an optimization algorithm. Friedman [16] used this observation to generalize AdaBoost to a class of algorithms called gradient boosting machines (GBM). In this general framework, a loss function $L(y, h(x))$ is chosen, and it is minimized using gradient descent. The model $h(x)$ is non-parametric and its structure is determined by the architecture chosen in the particular implementation of GBM. In most applications, GBM is built as an ensemble of base learners consisting of shallow trees.

The algorithm supports squared loss, absolute loss, t-distribution loss, log-loss, exponential loss, etc. Usually squared loss is used for continuous outcome, and log-loss is used for binary outcome. In minimizing the overall loss $\sum_{i=1}^{n} L(y_i, h(x_i))$, the prediction function $\hat{h}(x)$ is constructed in an additive way:

$$\hat{h}(x) = \hat{f}_0(x) + \sum_{m=1}^{M} \rho_m \hat{f}_m(x).$$

As shown in Figure 6, the algorithm works as follows:
- Take $\hat{f}_0(x)$ to be the baseline (e.g., overall mean or overall log-odds);
- In each stage $m$ $(m = 1, \ldots, M)$, update the prediction function in the direction $\hat{f}_m(x)$ where the total loss decreases;
- This direction is the negative gradient (gradient descent). Hence each base learner $f_m(x)$ is fit to the negative gradient of the loss function;
- The updated predictor is $h_{m-1}(x) = \hat{f}_0(x) + \rho_1 \hat{f}_1(x) + \cdots + \rho_{m-1} \hat{f}_{m-1}(x)$ where the $\rho_j$'s are step sizes or learning rates which are found using line search or set at a constant value;
- For squared error loss, the negative gradient is simply the error $\epsilon_{mi} = y_i - h_{m-1}(x)$ from previous stage; and

- For log-loss, the negative gradient is the error $\epsilon_{mi} = y_i - p_{m-1}(x_i)$, where $p_{m-1}(x) = \exp(h_{m-1}(x))/(1 + \exp(h_{m-1}(x)))$, is the predicted probability from stage $(m-1)$.

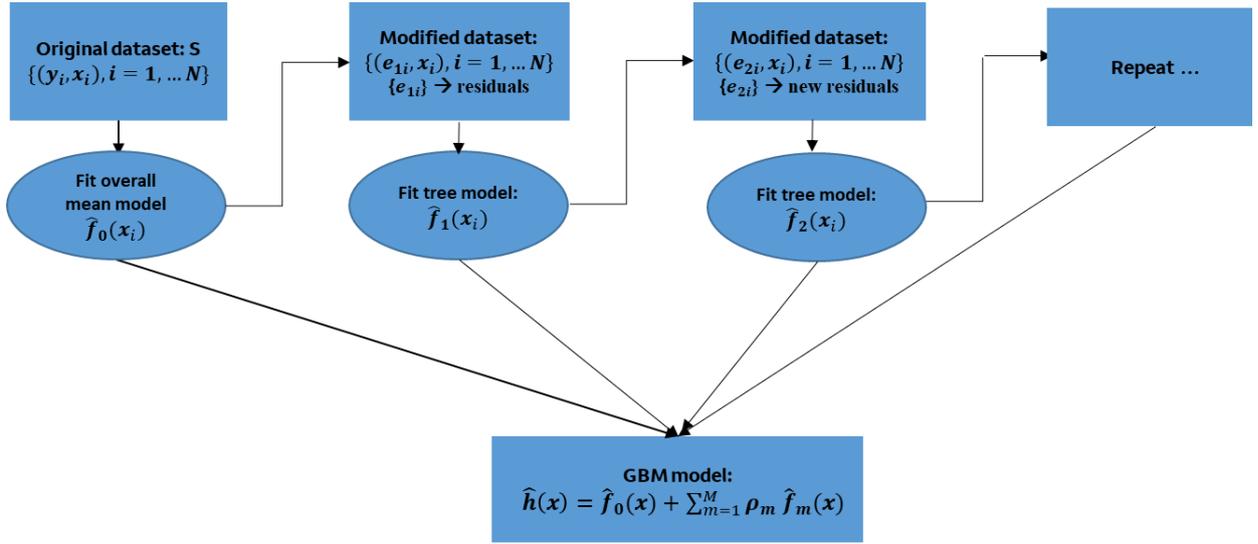

Figure 6. Illustration of GBM

A slight variation of GBM, proposed by Friedman [17], is stochastic gradient boosting which is motivated by the bagging method. At each stage, only a random subsample of the training data (drawn without replacement) is used to fit the tree. This can potentially reduce overfitting and produce a more robust model. Friedman [17] observed that a sampling rate between 0.5 and 0.8 gives good results.

XGBoost (extreme gradient boosting), by Chen and Guestrin [18], is a distributed, regularized gradient boosting algorithm that has become very popular. It minimizes

$$\sum_{i=1}^{n} L(y_i, h(x_i)) + \sum_{m=1}^{M} \Omega(f_m), \quad \text{where}$$

$$\Omega(f_m) = \gamma |f_m| + \frac{1}{2}\lambda \sum_{j=1}^{|f_m|} w_{mj}^2 + \alpha \sum_{j=1}^{|f_m|} |w_{mj}|,$$

where $|f_m|$ is the number of leaves in the tree. The first term $\gamma|f_m|$ in $\Omega(f_m)$ penalizes the number of leaves, and the $L_1$ and $L_2$ terms penalize the fitted values in each leaf. The trees are built in an additive way as in the original GBM algorithm.

The hyper-parameters associated with GBM include the previously discussed parameters in the tree algorithm plus two additional parameters: number of trees $M$ and learning rate $\rho_m$. Typically, shallow trees with small depth or number of nodes and a constant learning rate are used; small learning rates give better results [19]. For the number of trees, a small number under-fits the data while a large number over-fits the data. The best number of trees depends on the complexity of the underlying model, learning rate and complexity of each tree. XGBoost adds three additional hyper-parameters to control the strength of regularization. See Chen and Guestrin [18] for details.

### 2.2.4 Example (continued)

We return to the example in Section 2.2.2 to illustrate the results for XGBoost version of GBM. The green curve in Figure 5 is the ROC curve for GBM, and its AUC is 0.865, computed on test data. Like RF, GBM is better than logistic regression performance in this case, and is slightly better than RF. We have compared RF and GBM on a large number of datasets, and often GBM tends to perform slightly better. This is also consistent with what has been observed in the literature [19].

Figure 7 compares the predictive performance of logistic regression against GBM on hold-out data for the mortgage example discussed before. The algorithm was trained on data before 2005 and tested on data from 2005-2013. The x-axis shows time (in years) and y-axis shows prediction error after the actual values have been subtracted off. So the horizontal line at 0 corresponds to the actual data, red is the prediction error for logistic regression, and green is for GBM. We see that GBM also performs better overall for hold-out-testing.

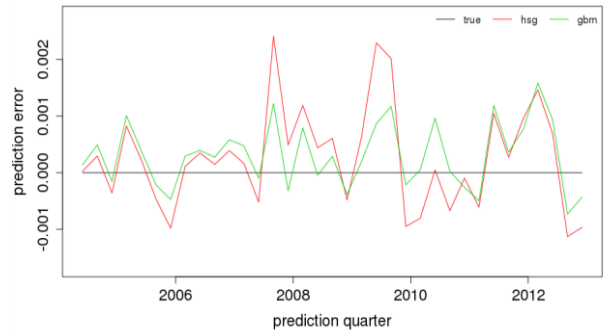

Figure 7: Comparison of predictive performance over time: test results on hold-out data

## 3 Artificial Neural Networks

Artificial neural networks (ANNs) were conceived as attempts to mimic biological neuronal networks. They are constructed by connecting the output of certain neurons to the input of other neurons, forming a directed, weighted graph, as in Figure 8.

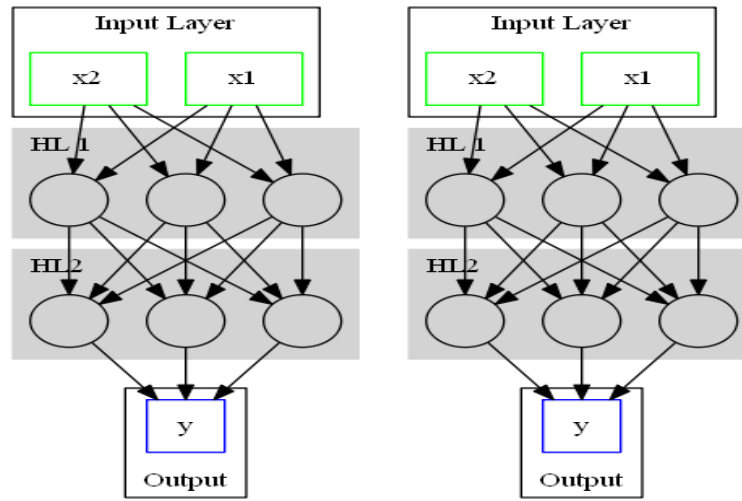

Figure 8: Example of FFNN

A feedforward neural network (FFNN) is an ANN wherein connections between the units are only in the forward direction. An FFNN usually has one input layer and one output layer. Without a hidden layer, it is called single-layer perceptron; with one or more hidden layers, it is called multi-layer perceptron, or MLP. The example in Figure 8 has two hidden layers. If the nodes in each layer are fully connected with the nodes from its previous layer as in Figure 8, the network is referred to as fully connected FFNN. It represents the most basic ANN structure. There has been extensive recent research in, and applications of, more complex network architectures, many of which go under the rubric of Deep Learning. These are beyond the scope of the current article, but see Nielsen [20] and references therein.

### 3.1 FFNN Architecture

A typical FFNN has several layers consisting of 1) input layer, 2) hidden layers, and 3) output layer. Each layer is made up of individual computational neurons denoted by ovals in Figure 8.

i) The input layer takes the input variables, as in any models/algorithms. These variables are generally numeric (coded to be numeric for non-numeric data). For example, categorical variables are often coded as a set of indicator variables. In Figure 8, the input layer is shown with two input variables.

ii) The hidden layer sits between input layer and output layer. Each computational layer of the neural network consists of one or more computational neurons (nodes). In Figure 8, there are two hidden layers of computational neurons. Each neuron is a simple unit that finds a linear combination of the inputs to that neuron, and then applies a (usually non-linear) activation function to the result. The network thus consists of connections, each connection transferring the output of a neuron $j$ in one layer to the input of a neuron $i$ in the next layer. In this sense $j$ is the predecessor of $i$, and $i$ is the successor of $j$. The linear combination consists of a weight $w_j$ multiplied by each input $x_j$. A constant called the bias ($b_j$) is added. The weights and bias are parameters "learned" in the training of the neural network:

$$z_i = \sum_{j=1}^{J} w_{ij} x_j + b_i.$$

The output (or activation) of the neuron is an additional function $g(\cdot)$ applied to that linear combination as $a_i = g(z_i)$. The choice of activation function can be made at the individual neuron level, but in practice are usually chosen for entire layers. Some commonly used activation functions are: i) sigmoid, $g(z) = \frac{1}{1+e^{-z}}$, ii) hyperbolic tangent, $g(z) = \tanh(z) = \frac{e^z - e^{-z}}{e^z + e^{-z}}$, iii) rectified linear units (ReLU), $g(z) = \max(0, z)$; or iv) identity $g(z) = z$. Each of the two hidden layers (HL1 and HL2) in Figure 8 has 3 hidden computational neurons. Since the network is fully connected, the output of each neuron on the layer is used as one of the inputs on the following layer.

iii) The output layer takes the output of the final hidden layer as input, and produces the predicted value of the network.

An FFNN may be adapted to different machine learning tasks by appropriately choosing the output layer. For example:

(1) A single node with an identity activation function represents a univariate regression task;
(2) A single node with a sigmoid activation function can be used for a binary classification, when the target takes values in binary events;
(3) A set of $k$ output nodes can be used for a $k$-class classification task using the "softmax" activation function. Each output node gives the probability that the corresponding observation belongs to one of the $k$ classes.

### 3.2 Training an FFNN

The weights (and bias) of each neuron are the unknown parameters in an FFNN that need to be learned from data. For estimation, we first define an appropriate loss function and then choose the weights and biases to minimize the loss function. Typically, we use squared error loss for continuous responses and cross entropy or log-loss for binary responses.

Denote the loss function as $L(\boldsymbol{\theta}|\boldsymbol{y},\boldsymbol{x})$ where $\boldsymbol{\theta}$ is the collection of all weights and biases. To train the weights and bias, a commonly used approach is gradient descent. However, computing the gradient $\nabla_{\boldsymbol{\theta}} L(\boldsymbol{\theta}|\boldsymbol{y},\boldsymbol{x})$ in neural networks can be challenging. Instead, the *back propagation algorithm* is used to calculate the gradient. First, all weights in the network are initialized. Then:

(1) Feed the data through the network, compute the output of each node based on the current weights;
(2) Compute the gradient of the loss function with respect to the last hidden layer;
(3) Work backwards through the network, computing the gradient of loss function with respect to the weights in $(j-1)$-th layer using the chain rule and the gradient with respect to the weights in $j$-th layer;
(4) Update the weights using gradient descent $\boldsymbol{\theta}_{i+1} = \boldsymbol{\theta}_i - \rho \nabla_{\boldsymbol{\theta}} L(\boldsymbol{\theta}_i|\boldsymbol{y},\boldsymbol{x})$ where $\rho$ is the learning rate, and return to the first step until $\nabla_{\boldsymbol{\theta}} L(\boldsymbol{\theta}_i|\boldsymbol{y},\boldsymbol{x})$ is very close to 0.

A variety of sophisticated methods have been developed to improve the learning in FFNNs, including stochastic gradient descent, RMSProp, Adadelta [21], and Adam [22]These algorithms improve learning by using: 1) momentum, to prevent the gradients from changing too rapidly/overcorrecting; and 2) adaptive learning rates, to balance speed with accuracy. In practice, the computation is done in batches of training data and the weights are updated based on a small subset of observations, called a mini-batch. This is referred to as "mini-batch learning".

Examples of hyper-parameters for NNs include the number of layers and number of nodes, as well as choice of activation functions. Many training algorithms have further hyper-parameters, e.g., learning rate. In general NNs have many more hyper-parameters than GBM and RF. Tuning all of these is extremely challenging and time-consuming if one aims at obtaining the best model performance with optimal parameter values in a large set of possible options. Techniques for hyper-parameter tuning is discussed in Section 4.

### 3.3 Deep Neural Networks

The terms `deep neural networks' (DNNs) and `deep learning' (DL) have become common in recent years. At a basic level, DNN can be just an FFNN with many hidden layers (deep), but there is no widely accepted number that indicates a transition from shallow networks to deep. Early papers used the term to refer to networks with as few as 3 hidden layers [23], while more recent usage deals with 11 to 19 hidden layers [24] or even more than 100 layers [25].

Perhaps a more interesting use of the term DNN deals with networks that have complex structure, with some of them including feedback as well as feedforward loops. Convolutional Neural Networks (CNNs) are among the most successful techniques used for image classification. While still feedforward in nature, the individual layers exploit parameter-sharing to reduce the total number of parameters in the network. If the data consist of 128 x 128 images, each node in the first hidden layer has to learn 128 x 128 weights (one for each pixel of the input). There are many such nodes in each layer, so training of the weights can be formidable. Instead, neurons in convolutional networks learn weights for a smaller filter that is applied successively to all regions of the input. So instead of 128 x 128 weights, each node learns only, for example, 16 x 16 weights. Successive layers use similar filters across subsets of the outputs of the previous layer's nodes. This reduction in the number of parameters allows deeper networks to be trained without increasing the computational demands for training. CNNs also have the desirable property that the filters are translation-invariant, meaning that they are able to recognize the same features regardless of location. For example, in an image analysis context, the network is able to recognize the image of an automobile: whether it is in the top left or bottom right of an image (see, for example, Chapter 9 of [26]).

Another important class of architectures, called Recurrent Neural Networks (RNNs), allows feedback loops: the output of a hidden layer can be an input to an earlier hidden layer. RNNs are often used to model sequence data, such as time series or sentences in language models. While there are many variants and configurations of RNNs, a particularly widespread and useful class is the Long Short-Term Memory (LSTM) network. It consists of complicated LSTM units that replace the usual artificial neurons. Each unit consists of an input node that functions as a usual artificial neuron, but adds a neuron to record the current state, as well as three gates: input, forget, and output. Each of the gates uses a combination of the input and the current state (via a recurrent loop) to update the input calculation, current state, or output, respectively. LSTM architecture has been particularly useful in capturing long range dependencies in sequence-type data (see, for example, Chapter 10, of [26])

### 3.4 Example

We use another dataset to illustrate the predictive performances of selected SML algorithms and compare them with logistic regression. This is a subset of 200,000 time series from a larger dataset dealing with various types of customer transactions. The response is binary: whether a particular event of interest (transaction) occurred or not. Subject matter expertise was used to summarize the time-series data into 75 hand-crafted predictors. We first fitted a logistic regression model with selected interactions using Lasso regularization. The RF and GBM algorithms were based on all 75 predictors. For the CNN algorithm, we used the original time series data as input rather than the hand-crafted features. The performances were compared using ten-fold cross validation: holding out $1/10^{th}$ of the data, training the model on the remaining, and testing the results on the $1/10^{th}$.

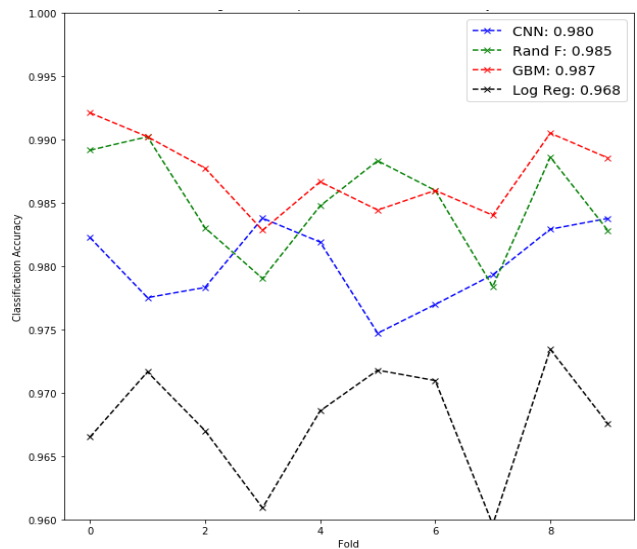

Figure 9: Comparison of the performances of logistic regression and three SML algorithms.
x-axis: ten different hold out cases; y-axis: AUC (classification accuracy).
Black lines: logistic regression results; blue denotes CNN; green denotes RF; and red denotes GBM.

Figure 9 shows the AUC results (classification accuracies) for ten different hold-out samples. The overall AUCs are given on the top right-hand area of the figure. All algorithms have high classification accuracies (AUCs). But the machine learning algorithms have slightly better performances: about a 2% improvement which can be substantial in practice. As noted earlier, GBM outperforms RF slightly. CNN has slightly worse performance but this might be due to our hyper-parameter tuning which was limited.

# 4 Hyper-parameter Optimization: General Discussion

The predictive performance of ML algorithms is highly dependent on the hyper-parameter setting, so choosing optimal values is critical. There is a large number of such parameters and possible configurations, so hyper-parameter tuning can be a hurdle when one does not have access to adequate computing resources. We describe below a few of the more common optimization approaches that are currently used.

## 4.1 Batch or non-sequential techniques

The idea in batch designs is to identify *upfront* a subset $s \in S$ of the hyper-parameter space, train the model at all the configurations in $s$, evaluate their performances on validation datasets, and select the best setting. Batch algorithms are not as efficient as the sequential ones that are discussed later, but they are used due to ease of implementation.

### 4.1.1 Grid search

This is the simplest and most naïve scheme. The left panel in Figure 10 shows a toy example with two hyper-parameters $p_1$ and $p_2$. Suppose our budget allows us to search over $N = 9$ model evaluations. We select a $3 \times 3$ grid in this approach, choosing three values for $p_1$ and three for $p_2$, then fitting and evaluating the model at all 9 combinations of these values. In Figure 10, the points for each hyper-parameter are equally spaced, but the selection of points can be done using subject-matter expertise. To describe grid-search more generally:

i) Use computational or other resource considerations to determine the total number $N$ of hyperparameter configurations to search over.
ii) Based on this constraint, determine the number of settings for each of the $K$ hyper-parameter, $M_k$, $k = 1, \dots, K$ with $N = M_1 \times \dots \times M_K$; for example, in Figure 10, $K = 2$, $M_1 = M_2 = 3$, and $N = 9$.
iii) Select the $M_k$ settings for each parameter. The cross-product of all these settings defines the subset $s \in S$, the grid consisting of all possible hyper-parameter configurations to be searched; again, Figure 10 is a simple example.
iv) Train the SML algorithm with these configurations, evaluate their model performances on the validation dataset, and select the best one.

Readers familiar with the field of statistical experimental design will recognize this scheme as an instance of a full factorial design. The advantages of this method are that it is easy to implement. However, a batch design can spend significant effort searching inefficient areas of the parameter space. In particular, if one of the hyper-parameters is not important (i.e., model performance does not depend on this parameter), resources would be wasted on training the ML algorithm at multiple values of this parameter while keeping others fixed. For example, in Figure 10, suppose parameter $p_2$ (y-axis) is not important. Then, by projecting onto the x-axis, we see that we have unnecessarily replicated each of the points $p_{1j}, j = 1,2,3$, three times. This point is well-known in computer experiments, and for this reason traditional design of experiments techniques such as full factorial are not used in these situations.

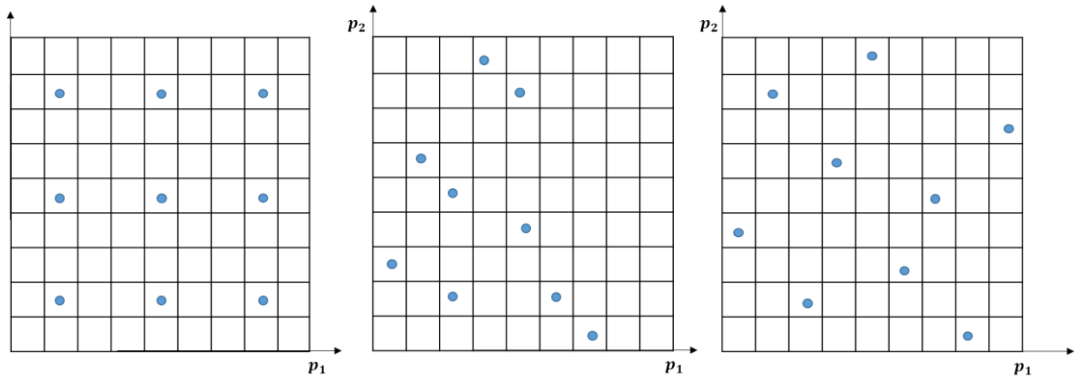

Figure 10: Visual comparison of: grid search (left); random search (middle); and Latin hypercube (right)

### 4.1.2 Random search

Random search algorithms are also easy to understand and use; see [27] for a discussion and comparisons in this context. To use a simple example, suppose we have the same budget of $N = 9$ as in the gird search case. The random search algorithm will select these nine points randomly (according to some mechanism) in the two-dimensional hyper-parameter space. Figure 10 illustrates a particular realization of this random selection. In general, we have a $K-$dimensional hyper-parameter space, and we have to randomly select $N$ configurations $\{(p_{1i}, p_{2i}, \ldots p_{Ki}), i = 1, \ldots N\}$ and then train and evaluate the algorithm's performance over these configurations. The most common way to do random selection is to take the marginal distributions to be uniform (in the original scale or a transformed scale such as log) and create a joint distribution as a product of the marginal distributions. Of course, if there is some a-priori information about the response surface, this can be used in the distribution.

One clear advantage with this scheme is that it allows one to select $N$ settings for each parameter and thus explore a wider area of the parameter space compared to grid search. If any particular parameter is not important (the fitted model performance does not vary with his parameter), one will not waste resources, as was the case with grid search. However, the selection of the configurations to search is purely left to chance, and it is possible that certain regions are not explored (see, for example, the top right corner in Figure 10). This can be a problem especially in high-dimensional spaces. The space-filling designs discussed next can alleviate this problem.

### 4.1.3 Space-filling designs

There are several other batch techniques in the computer experiments literature that could also be useful in this context. These have been mentioned but are not as widely used. In particular, space-filling designs are designed to spread out the points uniformly (in some sense) and fill out the design space. Examples include designs based on Latin hypercube sampling (LHS), max-min designs, and Sobol sequences (see the software package GPyOpt [https://github.com/SheffieldML/GPyOpt]; [28]; [29]). The rightmost panel in Figure 10 shows an LHS design with $N = 9$ points for two hyper-parameters. Note that each row and column has only one point, a property of LHS that holds more generally in higher dimensions. LHS will, in general, allocate the points to the search space more systematically than random search designs and hence cover it better. The class of max-pro LHS designs ( [28]) have very good projection properties in the sense of covering multiple lower-dimensional projections well.

## 4.2 Sequential Techniques

It is well known that sequential optimization algorithms are more efficient as they use information from previous exploration to identify good regions to search. There is an extensive literature on sequential algorithms, many of which are gradient based. In our context, gradients are not available or expensive to compute, so the focus has been on sequential search algorithms or use of surrogate models for the response surface $f(y|s)$, where $y$ denotes the response (model performance) as a function of $s$, the hyper-parameter configurations.

### 4.2.1 Hyperband

This technique, introduced in [30] is actually a hybrid batch-sequential algorithm; it is built on top of random search and a sequential technique called Successive-Halving or SH. Hyperband is most effective in situations where it is computationally expensive (many iterations and each is time consuming) to train an algorithm. Instead of iterating until convergence, hyperband trains the model for a small number of iterations, retains only a proportion of the top performing configurations based on this initial training, discards the rest, allocates more iterations to the retained ones, continues fitting, identifies and retains the top configurations, and keeps going until the best configuration is determined. Here, an iteration refers to a unit of resource allocated to the optimization algorithm, such as one epoch (a full pass over the dataset).

The hyper-band algorithm requires two input parameters: $R$, the maximum amount of resources, and $\eta$, a number that controls the proportion of configurations discarded in each SH round (this proportion is not typically ½ as suggested by the term "halving" in SH). These input parameters dictate the choices of other parameters in the algorithm: the batch size for random search, the number of iterations for the model training at each stage, and so on. The (limited) experimental studies in the original paper show that hyperband does well and outperforms several others including TPE discussed below. However, as we have noted, it is best suited for situations where model training is computationally intensive, as is the case with complex neural networks.

### 4.2.2 Sequential model-based global optimization (SMBO) techniques

These techniques rely on building a surrogate model and using it to optimize the sequential search process.
**a) Bayesian optimization with Gaussian Process** (BOGP), proposed in [31] uses a Gaussian process (GP) to model the response surface $[y|s]$, where $s$ represents the hyper-parameter setting and $y$ is the model performance on the validation dataset. (GP-based surrogate modeling has been used extensively in the design and analysis of computational experiments; see [29]). Specifically:
  i)   BOGP starts by treating $[y|s]$ as an unknown random function and uses a GP with zero mean and a suitable covariance kernel as its distribution $p(y|s)$;
  ii)  An initial batch sample $\{s_1, \ldots s_n\}$ of points are selected in the hyperparameter space;
  iii) The model is trained at these hyper-parameter configurations and their performances are computed;
  iv)  These values $z_1 = \{y_1, \ldots, y_n\}$ are treated as realizations from a multivariate normal distribution and the original GP is updated to get $p(y|s; z_1)$, another GP;
  v)   The value of an appropriate `acquisition function' is computed based on $p(y|s; z_1)$;
  vi)  The acquisition function is optimized to identify the next hyperparameter configuration to sample;
  vii) Steps iv) to vi) are repeated until a stopping criterion is met.

There are several choices for acquisition functions in step v) (see [31]), but the most common one is Expected Improvement criterion. There are different implementations of the basic idea in BOGP with a variety of GPs, acquisition functions, and other options (see [32] and the GPyOpt software package).

**b) Tree-structured Parzen estimator (TPE)** is another SMBO algorithm proposed and studied in [31]. Instead of modeling $p(y|s)$ as done in BOGP, TPE builds models for $p(s|y)$ and $p(y)$ and uses Bayes theorem to determine $p(y|s)$. TPE defines

$$p(s|y) = \begin{cases} l(s), & y < y^* \\ g(s), & y \geq y^* \end{cases}$$

where $y^*$ is the empirical $\gamma-$th quantile of the observed $y-$values (for a suitable value of $\gamma$). Recall that the goal is to select the configurations to minimize the predicted loss, so one wants to sample more from $l(s)$ compared to $g(s)$. Bergstra [31] showed that maximizing the expected information criterion is equivalent to maximizing $l(s)/g(s)$. Kernel-density estimators with Parzen windows are used to estimate the functions $l(x)$ and $g(s)$. The name 'tree-structured' comes from the fact that the hyper-parameters have a tree (or graph) since some of them are nested within others (such as number of nodes in number of layers). Paper [31] includes a small study that compares the performances of BOGP and TPE with random search and show that TPE performs the best.

### 4.2.3  Discussion

There are also other approaches in the literature for building a surrogate model, including the use of Random Forest. All sequential algorithms exploit previous information to do intelligent searches and hence are more efficient than batch techniques. The primary advantage of the latter is their simplicity. There are several open source software packages available (as of the writing of this paper) for implementing the sequential algorithms discussed above:

iv) **Spearmint:** https://github.com/HIPS/Spearmint
v) **RoBO:** http://automl.github.io/RoBO/
vi) **GPyOpt:** https://github.com/SheffieldML/GPyOpt
vii) **GPflowOpt:** https://github.com/GPflow/GPflowOpt
viii) **Hyperband:** https://github.com/zygmuntz/hyperband

## 5  Interpreting SML results

A major concern with SML algorithms is that the models are complex and the results are not easily interpretable. This is especially an issue in regulated sectors like banking where the results have to be explained to various stakeholders. In this section, we review several diagnostics that have been proposed in the literature and briefly summarize some recent results that have been developed.

But first, we digress briefly to make two points:
- There are no universally accepted definitions of the terms "interpretability" and "explainability". Additional ambiguity arises if we consider more terms such as transparency and model understanding. One reason for the confusion is the interchangeable use of these terms and the lack of precision in their common definitions. For example, interpretability is often viewed as the ability to explain or to provide meaning in understandable terms to a human. This is clearly subjective and can vary from person to person. The term explainability requires being able to understand the internal workings of the model and describe it to a decision maker. This may be possible in causal models or physical systems where there is subject matter knowledge. But it is not realistic to require that one should understand the inner-workings of predictive models based on observed data, especially when the variables are correlated and/or have complex interactions. In this paper, we eschew the differences among these terms and use them interchangeably for the most part. Readers interested in additional discussion are referred to [33] which provides perhaps best review of the terms and explanation of the differences.

- There seems to be a perception that traditional predictive models, such as linear and logistic regression, are easy to interpret. This is not so when the predictors have significant correlations. In this case, interaction terms such as $x_1 x_2$ cannot be separated from quadratic terms such as $x_1^2$ or $x_2^2$. Challenges in interpretation also arise when a model has many interactions or when we use complex semi- and non-parametric regression methods.

## 5.1 Global Diagnostics

### 5.1.1 Assessing variable importance

This technique provides an importance score for each predictor (or feature). It can be computed in different ways, and some depend on the particular SML algorithm. We describe the permutation-based approach which is agnostic to the type of ML algorithm used. The idea, originally proposed by Breiman [9] for random forests, is simple: a random permutation of the rows corresponding to a particular variable (column) will break the association of this variable with the response. So if the variable is important, the prediction accuracy should decrease significantly after permutation. The algorithm is implemented as follows:

1. Develop the predictive algorithm on the training data set;
2. For each variable $x_j$, randomly permute the rows of the variable while keeping the corresponding rows of other variables unchanged;
3. Obtain a new prediction using the permuted dataset; and
4. Compute the change in performance metric, such as AUC (binary outcome) or change in MSE (continuous outcome), after permutation. This gives the permutation importance score.

There are no thresholds or p-values associated with the importance scores. The typical approach is for the user to compute the variable importance scores for all predictors, select a top set and use subject matter knowledge or prior experience to determine if they make sense. Another way to use the variable importance analyses from SML algorithms is to determine if some of the variables are missing from a parametric regression model that is based on manual variable selection. One point, elaborated on in the section on PDPs below, is that the variable-importance results can be unreliable when there is high correlation among the predictors (features).

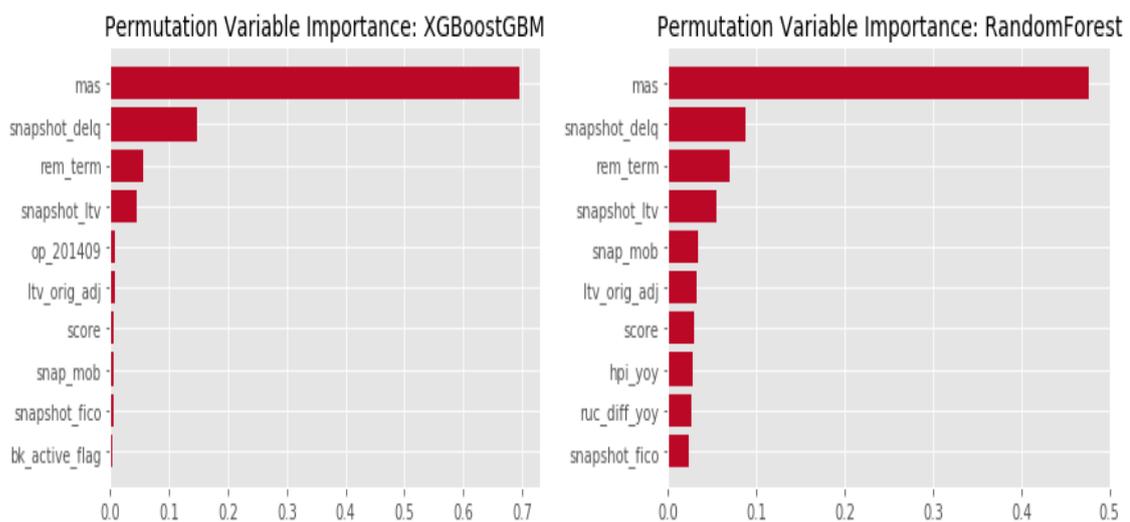

Figure 11. Permutation-based Variable Importance Plots

We use a subset of data on automobile loans to illustrate the results. There were 2.5 million accounts in this subset, and the response was binary: whether the loan had a particular type of default (charge-off, defined as 120 days past due) or not. Table 1 gives the list of predictors: macroeconomic variables, loan origination variables, and time-varying loan characteristics.

Table 1. Predictors for the automobile loan data

| Variable | Definition |
| --- | --- |
| mas | month after snapshot |
| snapshot_delq | delinquency cycles at "snapshot" (time at which prediction is made) |
| snap_mob | snapshot month on book |
| snapshot_fico | snapshot FICO |
| rem_term | remaining terms of the loan |
| snapshot_ltv | snapshot loan to value ratio |
| op_201409 | indicator for policy change in 2014 Q3 |
| PTI | origination payment to income ratio |
| ltv_orig_adj | origination loan to value ratio |
| score | customer score |
| hpi_yoy | house price index year over year change |
| ruc_diff_yoy | 12 month change of state unemployment rate |
| independent_dealer | indicator for independent dealer |
| MPV | indicator for multiple purpose vehicle |
| ZEINQ12MOS | number of inquiries in the last 12 months < 14 days |
| month | month indicators |

Figure 11 shows the variable importance analyses for the auto loan data. The left panel is for XGBoost and the right is for RF. The y-axis gives the variable names and x-axis gives value of the importance metric (normalized to add up to one across all the variables). Note that 8 of the top 10 variables picked by the two ML techniques are the same, and the top 4 are in the same order. Note also that the results in the two panels are qualitatively the same but differ in detail. This is a common occurrence as the two algorithms have different architecture and the results will not align completely. In particular, XGBoost tends to give most of its weight to the top few variables while RF tends to distribute it more evenly. The main reason is that RF selects a sample of the features (variables) at each split, so the key features are not present in all the trees/splits, hence diluting and redistributing their importance. Versions of XGBoost also have the option to subsample columns and will have similar behavior.

There are many other techniques that can be used to assess variable importance, including Sobol indices [34] for global sensitivity analysis, Global Shapely effects [35], derivative-based sensitivity [36], ANOVA decomposition based on ICE plots [37], etc.

### 5.1.2 Visualizing Input-Output Relationship: Partial Dependence Plots

Understanding the effect of the predictors in ML algorithms is not as straightforward as in parametric models. There are some techniques that have been proposed by Friedman [16]. Among the most important is the partial dependence plot (PDP), a graphical aid to interpret the effect of one or more predictors.

Let $X = (X_j, X_{-j})$, where $X_j$ is the covariates for which partial plot is to be generated, and $X_{-j}$ is the complementary subset of covariates. We have $\hat{f}(X) = \hat{f}(X_j, X_{-j})$ as the fitted model. Then the partial dependence function for $X_j = a$ is estimated by

$$\hat{f}_{par}(X_j = a) = \frac{1}{N}\sum_{i=1}^{N} \hat{f}(a, X_{-j,i}).$$

The partial-dependence plot (PDP) is obtained by computing $\hat{f}_{par}(X_j = a)$ for selected values of $a$ and plotting it. The PDPs proposed in Friedman [16] can be used for more than one-dimension, but in practice, only lower dimensional (1 or 2 dimensional) PDPs are easy to visualize. Moreover, higher-dimensional ones are computationally expensive. Since SML algorithms are non-parametric and can estimate inherent nonlinearity automatically, 1-D PDPs will exhibit the nonlinearity. Thus, they can also be used to assess the appropriateness of any variable transformation that have been used in parametric models. Similarly, 2-D PDPs can be used to examine higher-order relationships including the presence of any 2-d interaction effects.

The computation of PDPs can be expensive, so for continuous predictors, the values are usually computed only on a relatively small set of grid points in the $X_j$–space. For 1-d PDPs, a set of quantile points are chosen after trimming off the boundary (typically 1% data) to compute the PDPs. Similarly, for 2-d PDPs, PDP is computed on a selected 2-d grid.

Figure 12 shows examples of one-dimensional PDPs for *mas* (month after snapshot) for the auto loan data: one from XGBoost and one from RF – and. The y-axis is the response (log-odds in this case) and x-axis shows the values of the predictor (variable). The PDPs are computed and displayed at the quantiles of the predictors, given by the dots on the graph. The deciles are displayed as tick points along the x-axis to provide information about the distribution of the predictor. Note that there are more data points in the lower tails than upper tails, so the plot has more variability in the upper tail. While the PDPs in the two panels are qualitatively the same, there are also important differences indicating that the two non-parametric ML models are not totally aligned. This can happen, especially when the predictors are correlated.

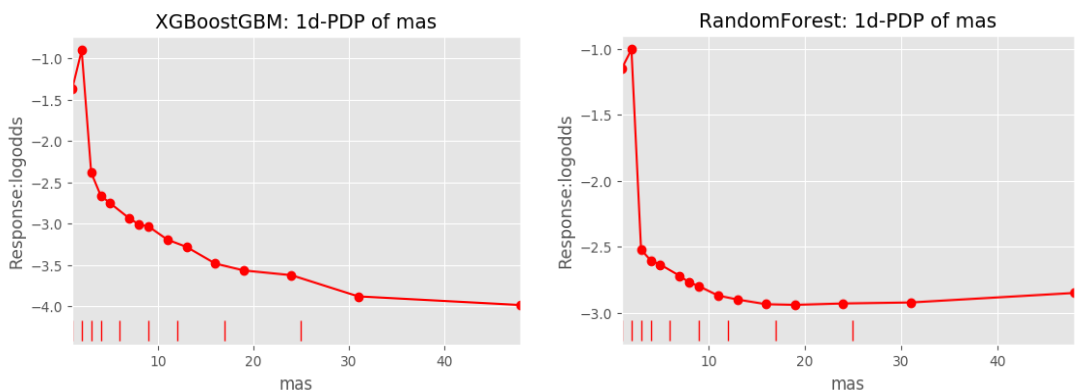

Figure 12. One-dimensional PDPs for the variable MAS – left panel: XGBoost; right panel: RF

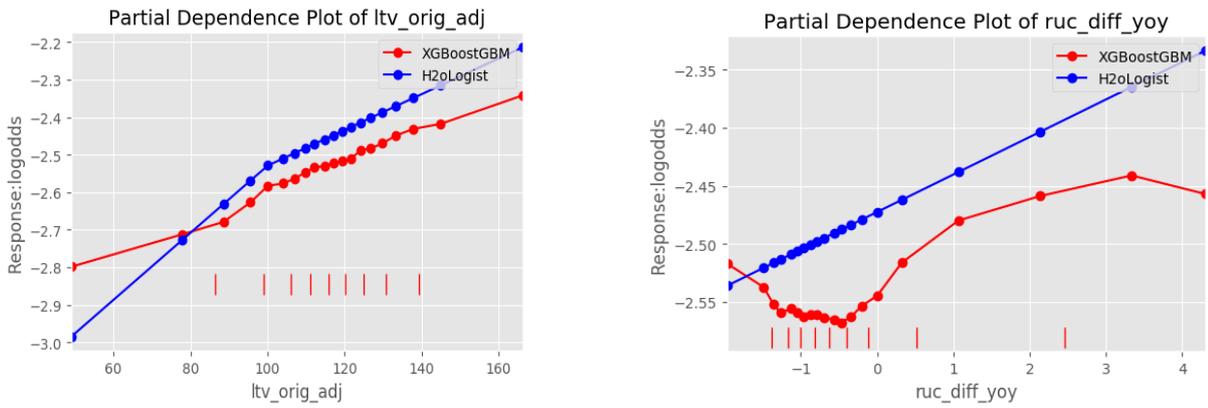

Figure 13. 1D-PDPs from XGBoost (red) overlaid with transformation used in Logistic Regression (blue)

The PDPs can also be used to assess the variable transformations used in parametric regression models. Figure 13 shows the comparison for two variables (for the same case study as before) by overlaying the 1D-PDP from XGBoost algorithm with the transformation used in the logistic regression model. In the left panel, the development team had used a piecewise linear transformation (blue) and the 1D-PDP (red) is reasonably close, suggesting that the choice by the original modelers was appropriate. In the right-hand panel, the development team fitted a linear relationship to the variable (blue). But the shape of the 1D-PDP (red) is somewhat different: decreasing initially and then increasing when the variable is greater than zero. Therefore, there is good reason to question fitting a linear relationship for this variable, as was done in the logistic regression model.

Figure 14 is one way of displaying a two-dimensional PDP. Each one of the curves is a 1-D PDP for mas (month after snapshot), but conditional on a fixed value of the second variable snapshot_delq (snapshot delinquency status), and the different curves show how the 1D-PDPs change as the values of snapshot_delq change. Parallel curves indicate no interaction. In this case, the two curves are largely parallel except for initial values of mas, indicating interaction at low values, but relatively little or no interaction at higher values of mas. From a business perspective, for snapshot delinquency status of 3 months (i.e., three months past due – light curve), the peak occurs at mas=1, i.e., it takes one more month to charge-off the loan. For accounts with snapshot delinquency status of 2 months (i.e., two months past due – dark curve), peak occurs at mas=2, i.e., it takes at least 2 additional months for charge off to occur.

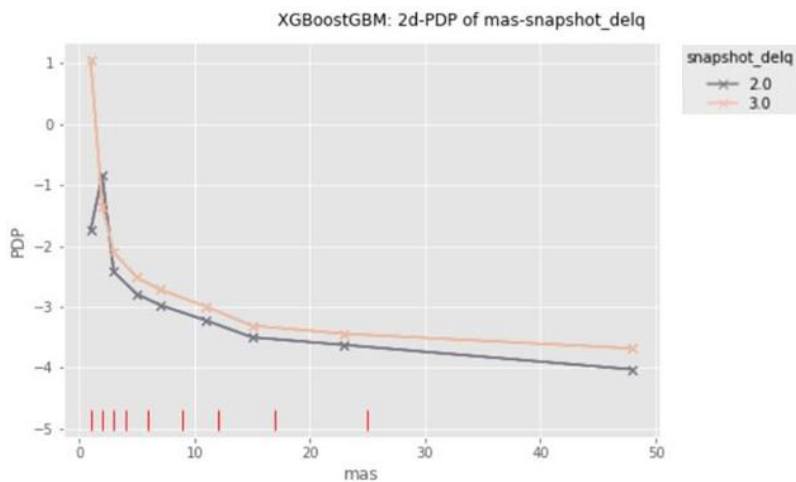

Figure 14. A two-dimensional PDP plot for snapshot_ltv vs mas

There are several potential limitations with the use of PDPs, some of which are due to the ML algorithms and others due to the diagnostic technique itself:

a) The plots can be quite jumpy (non-smooth) for RF and XGBoost, both of which are tree-based algorithms. This can sometimes mask the underlying relationships. PDPs from FFNNS will be smoother but they can over-smooth regions where there is a sharp regime change. These problems are to be expected with any nonparametric methods.

b) A considerable amount of caution should be exercised in interpreting the shape of the PDPs in regions where there is little data, typically in the tails.

c) One major difficulty with using PDPs arises when the predictors are even moderately correlated and as a consequence data are sparse in certain regions of the predictor space. The empirical calculations are based on permutations, so the PDPs include extrapolated points outside the region of observed data where the model fit can be highly questionable. So the results can be potentially unreliable. In practice, one should first check the correlation structure among the predictors before doing the analysis and also leverage Individual Conditional Expectation (ICE) plots [39] to identify or confirm the extrapolation of PDPs (see [38]). One could also consider removing selected variables among the highly-correlated pairs, or perform clustering to get rid of the high correlation before model refit. Other tools have been developed to ameliorate some of these problems: Accumulated Local Effects plots [40]; Accumulated Total Derivative Effects plots [41].

d) Since the ML algorithms are completely data-centric, the PDPs can violate subject matter knowledge, for instance it can be non-monotonic in situations where the relationship should be monotone.

## 5.2 H-statistics for Interactions

H-statistics are tools proposed in Friedman and Popescu [42] to quantify interaction effects among two variables and are quantitative summaries of the two-dimensional PDPs. For two variables $(x_j, x_k)$, the H-statistic is defined as

$$H_{jk} = \sqrt{H_{jk}^2},$$

where

$$H_{jk}^2 = \frac{\sum_{i=1}^N [f_{cpar}(x_{ij}, x_{ik}) - f_{cpar}(x_{ij}) - f_{cpar}(x_{ik})]^2}{\sum_{i=1}^N f_{cpar}^2(x_{ij}, x_{ik})}.$$

Here $f_{cpar}$ is the *centered* partial dependence function. This metric measures the proportion of variation in $f_{cpar}(x_{ij}, x_{ik})$ unexplained by an additive model. Its range is between 0 and 1, with a larger value indicating a stronger interaction pattern.

The above defined $H_{jk}$ is a relative (scaled) measure. If both the denominator and nominator shrink by half, $H_{jk}$ will not change. But when two variables are irrelevant, both denominator and nominator are small and $H_{jk}$ can be high due to instability. So the absolute H-statistics, which is just the numerator, is often used. This is given by

$$\widetilde{H}_{jk}^2 = \frac{1}{N} \sum_{i=1}^N [f_{cpar}(x_{ij}, x_{ik}) - f_{cpar}(x_{ij}) - f_{cpar}(x_{ik})]^2, \quad \widetilde{H}_{jk} = \sqrt{\widetilde{H}_{jk}^2}.$$

Calculation of H-statistics is also computationally expensive, so they are calculated on a grid as in 2-d PDPs.

H-statistics are quantitative single-number summaries of the 2d-PDPs but do not give as detailed information as the plots. Calibrating their value is a problem since their distributions are unknown and there are no critical values to determine when a computed H-statistic is large enough to indicate interaction. It is therefore recommended to use H-statistics along with the 2-d PDPs to see if the interaction effects make sense or not.

Table 2. H-statistics for the mortgage dataset

|              | mas   | snapshot_delq | rem_term | snapshot_ltv | op_201409 | ltv_orig_adj |
|--------------|-------|---------------|----------|--------------|-----------|--------------|
| mas          | NaN   | 0.535         | 0.248    | 0.419        | 0.097     | 0.043        |
| snapshot_delq| 0.535 | NaN           | 0.014    | 0.071        | 0.020     | 0.008        |
| rem_term     | 0.248 | 0.014         | NaN      | 0.043        | 0.038     | 0.029        |
| snapshot_ltv | 0.419 | 0.071         | 0.043    | NaN          | 0.029     | 0.050        |
| op_201409    | 0.097 | 0.020         | 0.038    | 0.029        | NaN       | 0.004        |
| ltv_orig_adj | 0.043 | 0.008         | 0.029    | 0.050        | 0.004     | NaN          |

Table 2 displays the H-statistics for six variables based on the XGBoost algorithm. The largest value of 0.535 corresponds to the variables in the 2D-PDP plot in Figure 14 where we saw strong interaction. The values corresponding to MAS vs rem_term and snapshot_ltv are also large. The other values are relatively smaller. In this case, one would use the 2D-PDPs for MAS vs the other two variables to understand the nature of the interactions.

H-statistics also share the limitations of PDPs discussed in the Section 5.1. In particular, they can be unreliable when the variables are highly correlated, due to the extrapolation issue mentioned earlier.

## 5.3 Local diagnostics

Local importance diagnostics explain the behavior of an individual prediction or the model in a local region. One application in banking deals with the requirement to provide reason codes for credit decisions. There are various local importance tools, such as LIME [43], Leave One Covariate Out (LOCO) [44], SHAP [45, 46], Quantitative Input Influence(QII) [47], etc. For neural networks, more methods are available, including Integrated Gradients [48], DeepLIFT [49], Layer-wise Relevance Propagation (LRP) [50], Derivative based Sensitivity Analysis [51], etc.

SHAP or SHapley Additive exPlanations [45, 46] is an approach that partitions the contributions to each predicted observation and allocates them to the different features. In a simple linear regression problem, consider the fitted model

$$\hat{y}_i = \hat{\beta}_0 + \hat{\beta}_1 x_{1i} + \hat{\beta}_2 x_{2i} + \cdots + \hat{\beta}_k x_{ki}.$$

The contributions of the $k$ different predictors to the fitted value $\hat{y}_i$ are just $\hat{\beta}_1 x_{1i}$, $\hat{\beta}_2 x_{2i}$, ... and $\hat{\beta}_k x_{ki}$. SHAP provides an analogous approach for complex ML algorithms. Paper [46] develops multiple ways to compute SHAP values including Kernel-SHAP and Tree-SHAP.

Locally interpretable model-agnostic explanations (LIME) is a technique developed in [43]. Given a point in the high-dimensional predictor space (typically one of the observations), LIME builds a simple local model around the point of interest and uses it to interpret the ML model locally. LIME is widely used, but it requires a local model to be built around each point of interest.

## 5.4 Surrogate models that are locally interpretable

Paper [52] develops a globally interpretable surrogate model for interpreting results from SML algorithms. (The notion of fitting surrogate models is common in many different application areas; see, for example, [53] for their use in computer experiments.) The technique in [52], called Surrogate Locally Interpretable model (SLIM), is based on recursive supervised partitioning. The idea is to approximate the fitted response surface of a complex ML algorithm by partitioning the input space using model-based regression trees and then fitting simple interpretable models locally (within the nodes). The splitting variables in the regression trees account for interactions while main-effects only models (GAM or spline-based main effects) are fit at each of the nodes.

## 5.5 Structured Neural Networks

There has been recent work on using neural network architecture to build intrinsically interpretable models. Paper [54] proposed explainable neural networks (xNNs) for fitting additive index models (AIMs):

$$f(\pmb{x}) = g_1(\pmb{\beta}_1^T \pmb{x}) + g_2(\pmb{\beta}_2^T \pmb{x}) + \ldots + g_K(\pmb{\beta}_K^T \pmb{x}),$$

where $\pmb{x}$ is a $P$-dimensional covariate, $\beta_k$ are projection indices, and $g_k(.), k = 1, \ldots, K$ are often called ridge functions. The $K = 1$ case is called a single index model [55]. Statistical readers will recognize AIM as projection pursuit regression [56]. Paper [54] used the structured neural network in the left panel of Figure 15 and gradient-based optimization techniques to fit AIM. The algorithm uses existing efficient techniques for fitting neural networks to estimate the parameters and fit the model. Generalized additive model network (GAMnet) can be viewed as special cases of xNN and can be used to fit the data to the following model (right panel of Figure 15):

$$f(\pmb{x}) = g_1(x_1) + g_2(x_2) + \ldots + g_P(x_P).$$

See [58] for most recent work on adaptive explainable neural network (AxNN) and related work. AxNNs are intended to perform the dual tasks of good predictive performance and explainable models.

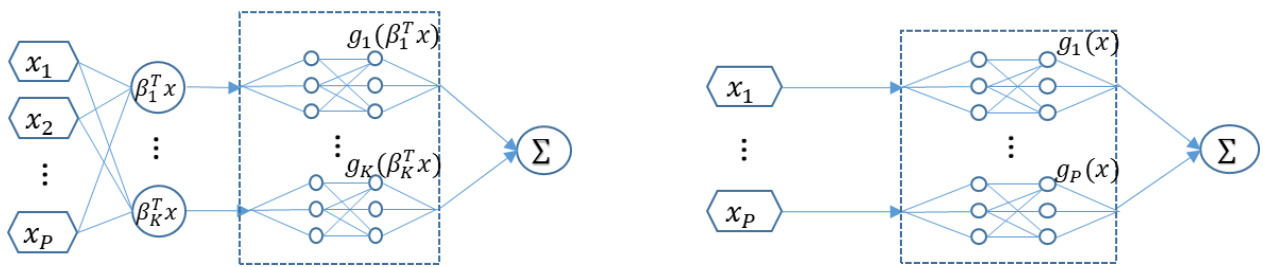

Figure 15: Illustration of structured NN architectures: xNN (left) and GAMnet (right)

## 6 Discussion

In general, RF, GBM, and FFNN are all competitive and none is universally better than another. Nevertheless, we provide some guidelines below based on our experience and what is known in the literature.

*i)* ***Computational speed:*** Neural networks take longer to fit. Even setting aside the larger number of hyper-parameters, in our experience, each NN fit seems to utilize fewer CPU cores compared to XGB or RF and hence takes longer time. We have compared RF and XGBoost quite extensively. While they both take about the same order of computational time, XGBoost appears to be slightly faster. Of course, any comparison will depend on the particular computing environment and implementation.

*ii)* **_Hyper-parameter tuning:_** Neural networks have many more hyper-parameters than RF and XGBoost, and tuning these parameters to find a good configuration is computationally very intensive. In our implementation of FFNNs, we tuned over only a limited number and range of hyper-parameters.

*iii)* **_Predictive performance:_** There is strong evidence in the literature, and confirmed by our own experience, that XGBoost has better predictive performance than RF. But the difference appears to be small in most cases. We have not tested the predictive performance of FFNNs extensively because of the challenges with hyperparameter optimization. We know from results in the literature that complex NNs used in Deep Learning yield excellent performance in pattern recognition problems. The closest analogy to pattern recognition in our applications is time-series data and panel data.

*iv)* **_Smoothness of fitted surfaces_:** Random forests and XGBoost are based on trees, so the fitted response surfaces are generally not smooth. Even though they are based on "averaging" of trees, they can still be non-smooth, especially with small datasets. As the sample size gets very large, the surfaces get smoother. On the other hand, neural networks (NNs) are based on linear combinations of the predictors, so the response surfaces are quite smooth. NNs also yield derivatives of the fitted responses as a by-product of the algorithm. These derivatives are useful in diagnostics.

One of the challenges with most SML algorithms is the difficulty in accommodating model constraints such as monotonicity of the input-output relationships. There is, however, research and new developments in this directions, especially with the use of NNs. There are many different software implementations of ML algorithms, with differences in how the same class of algorithm (such as GBM) is developed and implemented, differences in how they handle missing data, deal with categorical predictors, and so on. This is a fast changing landscape, and users have to be cognizant in keeping track of new developments.

## 7 Concluding Remarks

We have provided an overview of SML techniques and have illustrated their usefulness on applications to banking. While the techniques have now been around for some time, their widespread use is still somewhat limited, and there is still much to be learned about their performance in different situations. One frequently asked question is which algorithm to use and when. There are no clear-cut answers to questions like this, just like there is no straightforward answers to questions in modeling and data analysis, such as which variable selection technique or missing data analysis technique to use, what is the best goodness-of-fit-test, etc. Users have to gain experience and develop their own insights over time.

## 8 Acknowledgments

The authors are grateful to Xiaoyu Liu for her contributions to this work while she was at Wells Fargo. We thank Soroush Aramideh, Fernando Cela Diaz, and Rahul Singh for their comments on hyper-parameter optimization.

**References**

[1] L. Breiman, "Statistical Modeling: The Two Cultures (with comments and a rejoinder by the author)," *Statistical Science,* vol. 16, pp. 199-231, 2001.

[2] G. V. Kass, "An Exploratory Technique for Investigating Large Quantities of Categorical Data," *Applied Statistics,* vol. 29, no. 2, pp. 119-127, 1980.

[3] R. Quinlan, C4.5: Programs for Machine Learning, Morgan Kaufmann Publishers, 1993.

[4] L. Breiman, J. H. Friedman, R. A. Olshen and C. J. Stone, Classification and Regression Trees, Belmont: Wadsworth, 1984.

[5] T. Hothorn, K. Hornik and A. Zeileis, "Unbiased recursive partitioning: A conditional inference framework," *Journal of Computational and Graphical statistics,* vol. 15, no. 3, pp. 651--674, 2006.

[6] T. Hastie, R. Tibshirani and J. Friedman, Elements of Statistical Learning: Data Mining, Inference and Prediction (second edition), New York: Springer-Verlag, 2009.

[7] L. Breiman, "Bagging Predictors," Department of Statistics, University of California, Berkeley, 1994.

[8] L. Breiman, "Bagging Predictors," *Machine Learning,* vol. 24, pp. 123-140, 1996.

[9] L. Breiman, "Random Forests," *Machine Learning,* vol. 45, pp. 5-32, 2001.

[10] M. Kearns, "Thoughts on Hypothesis Boosting," 1988.

[11] M. Kearns and L. Valiant, "Crytographic Limitations on Learning Boolean Formulae and Finite Automata," in *Proceedings of the Twenty-First Annual ACM Symposium on Theory of Computing*, Seattle, 1989.

[12] R. E. Schapire, "The Strength of Weak Learnability," *Machine Learning,* vol. 5, pp. 197-227, 1990.

[13] Y. Freund and R. E. Schapire, "A Decision-Theoretic Generalization of On-Line Learning and An Application to Boosting," 1995.

[14] R. E. Schapire and Y. Freund, Boosting: Foundations and Algorithms, MIT Press, 2012.

[15] L. Breiman, "Arcing the Edge," Department of Statistics, University of California, Berkeley, 1997.

[16] J. H. Friedman, "Greedy Function Approximation: A Gradient Boosting Machine," *The Annals of Statistics,* vol. 29, pp. 1189-1232, 2001.

[17] J. H. Friedman, "Stochastic Gradient Boosting," *Computational Statistics & Data Analysis,* vol. 38, pp. 367-378, 2002.

[18] T. Chen and C. Guestrin, "XGBoost: A Scalable Tree Boosting System," in *Proceedings of the 22nd ACM SIGKDD International Conference on Knowledge Discovery and Data Mining*, San Francisco, 2016.

[19] A. N.-M. Rich Caruana, "An Empirical Comparison of Supervised Learning Algorithms Using Different Performance Metrics," in *ICML06*, Pittsburgh, 2005.

[20] M. Nielsen, Neural Networks and Deep Learning, 2017.

[21]   M. D. Zeiler, "ADADELTA: An Adaptive Learning Rate Method," *arXiv,* p. 1212.5701, 2012.

[22]   D. P. Kingma and J. Ba, "Adam: A Method for Stochastic Optimization," *arXiv,* p. 1412.6980, 2014.

[23]   G. E. Hinton, S. Osindero and Y.-W. Teh, "A Fast Learning Algorithm for Deep Belief Nets," *Neural Comput,* vol. 18, no. 7, pp. 1527-1554, 2006.

[24]   K. Simonyan and A. Zisserman, "Very Deep Convolutional Networks for Large-Scale Image Recognition," *arXiv,* p. 1409.1556, 2014.

[25]   K. He, X. Zhang, S. Ren and J. Sun, "Deep Residual Learning for Image Recognition," *arXiv,* p. 512.03385, 2015.

[26]   I. Goodfellow, Y. Bengio and A. Courville, Deep Learning, Cambridge, MA: MIT Press, 2015.

[27]   J. Bergstra and Y. Bengio, "Random Search for Hyper-Parameter Optimization," *Journal of Machine Learning Research,* vol. 13, 2012.

[28]   V. R. Joseph, "Space-filling designs for computer experiments: A review," *Quality Engineering,* vol. 28, no. 1, pp. 28-35, 2016.

[29]   R. B. Gramacy, Surrogates: Gaussian Process Modeling, Design, and Optimization for the Applied Sciences, Chapman & Hall/CRC, 2020.

[30]   L. Li, K. Jamieson, G. DeSalvo, A. Rostamizadeh and A. Talwalkar, "Hyperband: A Novel Bandit-Based Approach to Hyperparameter Optimization," *Journal of Machine Learning Research,* vol. 18, no. 1, pp. 1-52, 2018.

[31]   J. Bergstra, R. Bardenet, Y. Bengio and B. Kégl, "Algorithms for hyper-parameter optimization," in *NIPS'11: Proceedings of the 24th International Conference on Neural Information Processing Systems*, Granada, Spain, 2011.

[32]   J. Snoek, H. Larochelle and R. P. Adams, "Practical Bayesian optimization of machine learning algorithms," in *NIPS'12: Proceedings of the 25th International Conference on Neural Information Processing Systems*, 2012.

[33]   A. B. Arrieta, N. Díaz-Rodríguez, J. Del Ser, A. Bennetot, S. Tabik, A. Barbado, S. García, S. Gil-López, D. Molina, R. Benjamins and others, "Explainable Artificial Intelligence (XAI): Concepts, taxonomies, opportunities and challenges toward responsible AI," *Information Fusion,* vol. 58, pp. 82-115, 2020.

[34]   I. Sobol, "Global sensitivity indices for nonlinear mathematical models and their Monte Carlo estimates," *Mathematics and Computers in Simulation,* vol. 55, pp. 271-280, 2001.

[35]   E. Song, B. L. Nelson and J. Staum, "Shapley Effects for Global Sensitivity Analysis: Theory and Computation," *J. Uncertainty Quantification,* vol. 4, no. 1, pp. 1060-1083, 2016.

[36]   S. Kucherenko and others, "A new derivative based importance criterion for groups of variables and its link with the global sensitivity indices," *Computer Physics Communications,* vol. 181, pp. 1212-1217, 2010.

[37]   J. Chen, L. Hu, V. Nair and S. Agus, "Formulation of individual conditional expectation (ICE) plot and the link with Sobol Indices," *Wells Fargo Internal Report,* 2018.

[38]   A. Goldstein, A. Kapelner, J. Bleich and E. Pitkin, "Peeking Inside the Black Box: Visualizing Statistical Learning with Plots of Individual Conditional Expectation," *eprint arXiv:1309.6392,* 2013.

[39] A. Goldstein, A. Kapelner, J. Bleich and E. Pitkin, "Peeking inside the black box: Visualizing statistical learning with plots of individual conditional expectation," *Journal of Computational and Graphical Statistics,* vol. 24, pp. 44-65, 2015.

[40] D. W. Apley, "Visualizing the effects of predictor variables in black box supervised learning models," *arXiv preprint arXiv:1612.08468,* 2016.

[41] X. Liu, J. Chen, J. Vaughan, V. Nair and A. Sudjianto, "Model interpretation: A unified derivative-based framework for nonparametric regression and supervised machine learning," *arXiv preprint arXiv:1808.07216,* 2018.

[42] J. H. Friedman and B. E. Popescu, "Predictive Learning via Rule Ensembles," *The Annals of Applied Statistics,* vol. 2, pp. 916-954, 2008.

[43] M. T. Ribeiro, S. Singh and C. Guestrin, "" Why should i trust you?" Explaining the predictions of any classifier," in *Proceedings of the 22nd ACM SIGKDD international conference on knowledge discovery and data mining*, 2016.

[44] J. Lei, M. G'Sell, A. Rinaldo, R. J. Tibshirani and L. Wasserman, "Distribution-free predictive inference for regression," *Journal of the American Statistical Association,* vol. 113, no. 523, pp. 1094--1111, 2018.

[45] S. M. Lundberg and S.-I. Lee, "A unified approach to interpreting model predictions," in *Advances in neural information processing systems*, 2017.

[46] S. M. Lundberg, G. G. Erion and S.-I. Lee, "Consistent individualized feature attribution for tree ensembles," *arXiv preprint arXiv:1802.03888,* 2018.

[47] A. Datta, S. Sen and Y. Zick, "Algorithmic transparency via quantitative input influence: Theory and experiments with learning systems," in *2016 IEEE symposium on security and privacy (SP)*, 2016.

[48] M. Sundararajan, A. Taly and Q. Yan, "Axiomatic attribution for deep networks," in *Proceedings of the 34th International Conference on Machine Learning-Volume 70*, 2017.

[49] A. Shrikumar, P. Greenside and A. Kundaje, "Learning important features through propagating activation differences," in *Proceedings of the 34th International Conference on Machine Learning-Volume 70*, 2017.

[50] A. Binder, G. Montavon, S. Lapuschkin, K.-R. Müller and W. Samek, "Layer-wise relevance propagation for neural networks with local renormalization layers," in *International Conference on Artificial Neural Networks*, 2016.

[51] X. Liu, J. Chen, V. Nair and A. Sudjianto, "A. Interpreting Supervised Machine Learning Algorithms with Derivative-based Tools: Case Studies," *Wells Fargo Internal Report,* 2019.

[52] L. Hu, J. Chen, V. N. Nair and A. Sudjianto, "Surrogate Locally-Interpretable Models".

[53] L. S. Bastos and A. O'Hagan, "Diagnostics for Gaussian process emulators," *Technometrics,* vol. 51, pp. 425-438, 2009.

[54] J. Vaughan, A. Sudjianto, E. Brahimi, J. Chen and V. N. Nair, "Explainable Neural Networks based on Additive Index Models," *The RMA Journal,* 2018.

[55] R. Dudeja and D. Hsu, "Learning Single-Index Models in Gaussian Space," 2018.

[56] J. H. Friedman and W. Stuetzle, "Projection pursuit regression," *Journal of the American statistical Association,* vol. 76, no. 376, pp. 817-823, 1981.

[57]     S. Kucherenko, S. Tarantola and P. Annoni, "Estimation of global sensitivity indices for models with dependent variables," *Computer physics communications,* vol. 183, pp. 937-946, 2012.

[58]     E. Song, B. L. Nelson and J. Staum, "Shapley effects for global sensitivity analysis: Theory and computation," *SIAM/ASA Journal on Uncertainty Quantification,* vol. 4, pp. 1060-1083, 2016.

[59]     J. Chen, J. Vaughan, V. Nair and A. Sudjianto, "Adaptive Explainable Neural Networks (AxNNs)," *arXiv preprint arXiv:2004.02353,* 2020.